\documentclass[
 amsmath,amssymb,
 aps,
prb, twocolumn,
]{revtex4-2}
\usepackage{xcolor}
\usepackage{graphicx}
\usepackage{dcolumn}
\usepackage{bm}
\usepackage[colorlinks=true,linkcolor=blue,allcolors=blue]{hyperref}%
\usepackage{subfiles}
\usepackage{tabularx}

\begin{document}

\title{Antiferromagnetism and large magnetoresistance in GdBi single crystal}
\author{Gourav Dwari, Souvik Sasmal, Bishal Maity, Vikas Saini, Ruta Kulkarni, Arumugam Thamizhavel}
\affiliation{Department of Condensed Matter Physics and Materials Science, Tata Institute of Fundamental Research, Homi Bhabha Road, Colaba, Mumbai 400 005, India}
\date{\today}

\begin{abstract}
	
Single crystal of the binary equi-atomic compound GdBi crystallizing in the rock salt type cubic crystal structure with the space group $Fm\bar{3}m$ has been grown by flux method. The electrical and magnetic measurements have been performed on well oriented single crystals.  The antiferromagnetic ordering of the Gd moments is confirmed at $T_{\rm N} = 27.5$~K.  The magnetization measurement performed at $2$~K along the principal crystallographic direction [100] did not show any metamagnetic transition and no  sign of saturation up to $7$~T. Zero field electrical resistivity reveals a sharp drop at $27.5$~K suggesting a reduction in the spin disorder scattering due to the antiferromagnetic alignment of the Gd moments.  The residual resistivity at $2$~K  is 390~n$\Omega$cm  suggesting a good quality of the grown crystal.  The magneto resistance attains a value of $1.0~\times~10^{4}\%$ with no sign of saturation, in a field of $14$~T, at $T = 2$~K.  Shubnikov de Hass (SdH) oscillations have been observed in the high field range of the magnetoresistance with five different frequencies corresponding to the extremal areas of the Fermi surface.  Analysis of the Hall data revealed a near compensation of the charge carriers accounting for the extremely large magnetoresistance.

\end{abstract}

\maketitle
\section{Introduction}

The recent focus in condensed matter physics is on the observation of a new state of topological quantum matter in novel materials that have interesting band structures~\cite{lv2021experimental, vergniory2019complete, narang2021topology}.  Dirac and Weyl semimetals that have gained importance these days  fall in this category.  The Dirac point has a linear dispersion and can also be considered as a pair of Weyl points in $k$-space that are protected by both crystalline inversion $\mathcal{I}$  and time reversal $\mathcal{T}$ symmetries.  When either of the symmetries is broken the Dirac semimetal evolves into a Weyl semimetal~\cite{hu2019transport, armitage2018weyl}.  The prototypic Dirac semimetals Cd$_3$As$_2$ and Na$_3$Bi have been theoretically predicted as Dirac semimetals and later realized in experiments~\cite{PhysRevB.88.125427, liu2014stable, PhysRevB.85.195320, liu2014discovery}.  One of the interesting features of these Dirac semimetals is that they exhibit extremely large magnetoresistance,  ultra-high mobility and depict chiral anomaly, quantum Hall effect etc., due to the symmetry protected band crossings.  The extremely large magnetoresistance (XMR)  has been observed in several binary intermetallic compounds like WTe$_2$, MoSi$_2$, WSi$_2$, NbP, MoP$_2$, WP$_2$ etc.  Ultra-high mobility ($\approx 10^4$ cm$^2$/V s) and electron-hole resonance  with relatively lesser carrier concentration are the reasons for the XMR in these compounds~\cite{Ali_2014, PhysRevB.97.205130, PhysRevB.102.115158, shekhar2015extremely, kumar2017extremely}.  Although, such XMR is observed in non-magnetic intermetallic compounds, the natural extension of these studies is to combine the topological aspects with the strong electronic correlations.  The strong electronic correlation is observed in $f$ electron systems and the rare-earth monopnictides $RX$, where $R$ is a rare-earth element and $X$ is Sb or Bi, crystallizing in the simple rock-salt type cubic crystal structure was the default choice.   To start with, the non-magnetic LaBi and LaSb have been studied and both the compounds were showing XMR of the order of $10^5\%$~\cite{tafti2016temperature, tafti2016resistivity}.  The field dependence of the electrical resistivity  of these two compounds showed an upturn and plateau region which were observed in several semimetallic compounds and hence led to the construction of a universal triangular phase diagram~\cite{tafti2016temperature, PhysRevB.102.115158}.  Several of the RBi compounds like PrBi, SmBi,  ErBi, HoBi etc., have been reported recently also exhibit large MR~\cite{PhysRevB.99.245131, PhysRevMaterials.5.054201, PhysRevB.102.104417, Wu_2019}. It is interesting to note that in spite of the increasing $f$ electron count the XMR is still observed which reveals that the band structure may remain unchanged due to the highly localized nature of the $f$ electrons.   PrBi does not show any magnetic ordering due to the singlet ground state of the crystal electric field split  $(2J+1) = 9$-fold degenerate of Pr$^{3+}$ ion~\cite{PhysRevB.99.245131}.  The higher rare-earths HoBi and ErBi show magnetic ordering at 5.9~K and 3.6~K, respectively~\cite{ Wu_2019, PhysRevB.102.104417}.  Furthermore, Li et al~\cite{li2017predicted} predicted GdBi as a compensated semimetal with non-trivial band topology in its antiferromagnetic state.  

In this work, we present a systematic investigation on the transport and magnetic properties of GdBi single crystal.  The magnetic measurement depict a long range antiferromagnetic ordering of Gd$^{3+}$ moments at $T_{\rm N} = 27.5$~K.   The magnetic ordering is further confirmed from the electrical resistivity and specific heat studies.  The electrical resistivity displayed a typical semimetallic character in zero field measurements.  With the application of magnetic field the resistivity revealed an upturn at low temperature similar to a metal-insulator-like transition. At sufficiently high fields in the range $10-14$~T, the quantum oscillations are observed in magnetoresistance.  Five different frequencies have been observed in Shubnikov-de Haas (SdH) oscillations.

\section{Experimental Methods}

Although, GdBi melts congruently at $1770~^{\circ}$C~\cite{okamoto1990binary}, the high melting temperature precludes its growth by Czochralski method due to the high vapour pressure of Bi at such high temperature.  Hence, the single crystal of GdBi has been grown by self flux method using molten Bi as flux.  High purity Gd ingots (99.9\%, Alfa Aesar) and Bi lumps (99.998\%, Alpha Aesar) were packed into a baked round bottomed  alumina crucible in the  molar ratio of Gd:Bi = 20:80 and sealed under vacuum in a quartz tube.  The quartz tube was subsequently placed in a box-type resistive heating furnace and the sample was heated to 1100~$^{\circ} $C with a heating rate of 50~$^{\circ}$C/h and held at this temperature for about 12 h for homogenization.  Then the furnace was cooled down to 920~$^{\circ}$C at a rate of 1~$^{\circ}$C/h followed by 3 days of annealing.  We centrifuged the excess Bi-flux at $920$~$^{\circ}$C in order to avoid the formation of GdBi$_2$ which crystallizes below 910~$^{\circ}$C.  Cubic single crystals of GdBi with typical dimensions of $\approx$ 2 $\times$2 $\times$ 2~mm$^3$ were obtained. The compositional analysis of the grown crystals was performed using Energy Dispersive X-ray Spectroscopy (EDX).  The phase purity of the crystals were confirmed from   x-ray diffraction (XRD) performed in a PANalytical x-ray diffractometer equipped with a monochromatic Cu-$K_{\alpha}$ x-ray source ($\lambda = 1.5406$~\AA) and the crystals were oriented along the principal crystallographic direction using Laue diffraction using a polychromatic x-ray source.  The crystals were cut into desired shapes using a spark erosion electric discharge machine (EDM).  Magnetic measurements were performed in a SQUID magnetometer (MPMS, Quantum Design, USA) and the electrical and heat capacity measurements were performed in a physical property measurement system (PPMS, Quantum Design, USA).

\section{Results and Discussion}

\subsection{X-ray diffraction}

\begin{figure}[!]
	\includegraphics[width=1\linewidth]{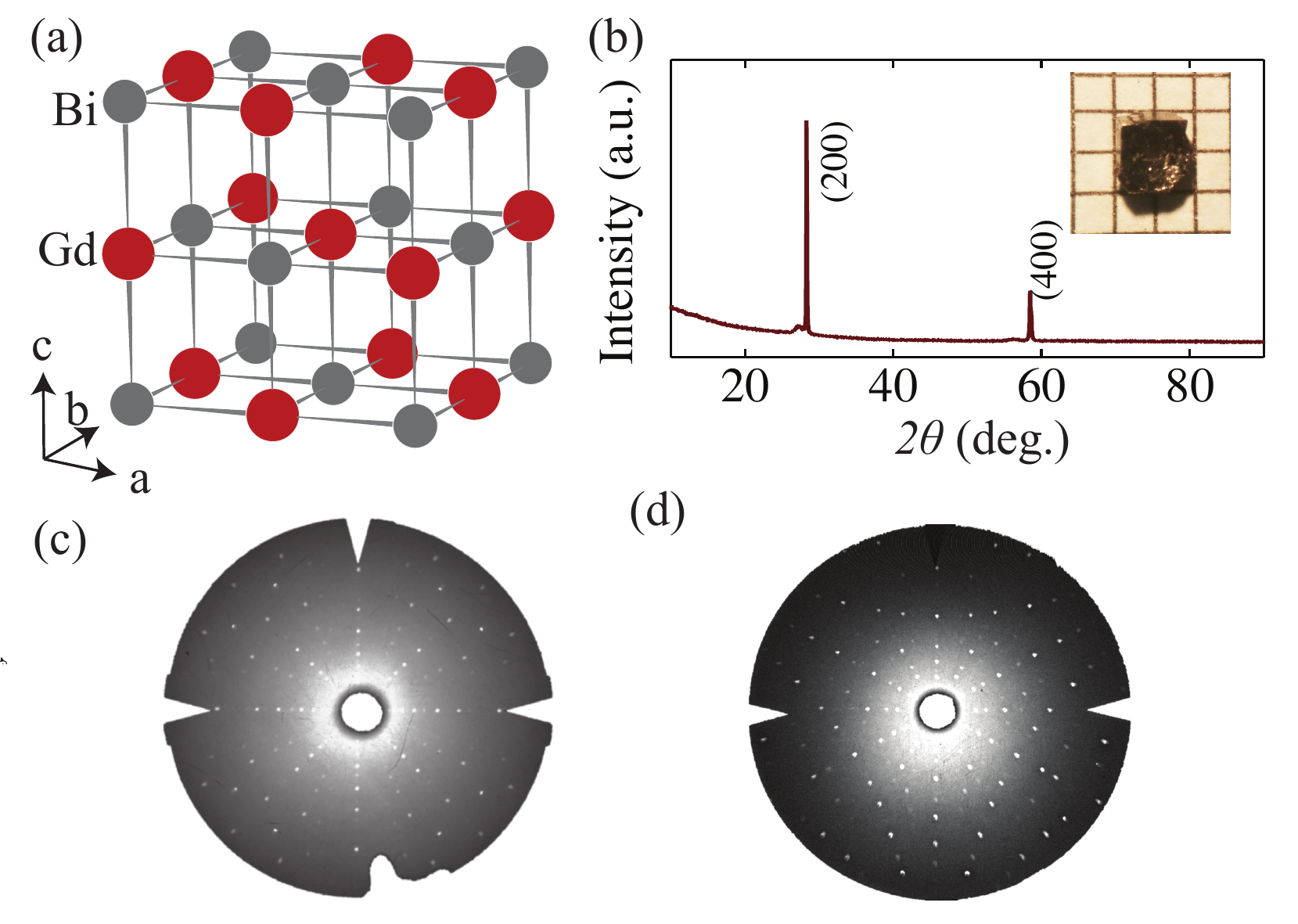}
	\caption{\label{Fig1} (a) Crystal structure of GdBi. (b) Room temperature XRD pattern of a single crystal of GdBi, (inset) as grown single crystal of GdBi. Laue pattern of GdBi for (c) (100) (or equivalent) plane and (d) (111) (or equivalent) plane.}
\end{figure}

The crystal structure of GdBi  is shown in Fig.~\ref{Fig1}(a). A small piece of the as grown crystal was subjected to XRD at 300~K with $2\theta$ scan ranging from 10 to 90$^{\circ}$, peaks corresponding to $(h00)$-planes are observed at  Bragg angles 28.2$^{\circ}$ and 58.44$^{\circ}$ thus confirming the flat plane of the crystal  to be (100)-plane. According to the previous report, GdBi crystallizes in NaCl type structure with space group \textit{Fm-3m} (No. 225)~\cite{yoshihara1975rare}.      The Laue pattern corresponding to (100) and (111) planes are shown Fig.~\ref{Fig1}(c) and (d).  Well defined circular spots together with the four fold symmetry in (100)-plane confirmed the good quality of  single crystal. The composition of grown crystal was confirmed from EDX measurement.

\subsection{ Magnetic properties} 

\begin{figure}[!]
	\includegraphics[width=1\linewidth]{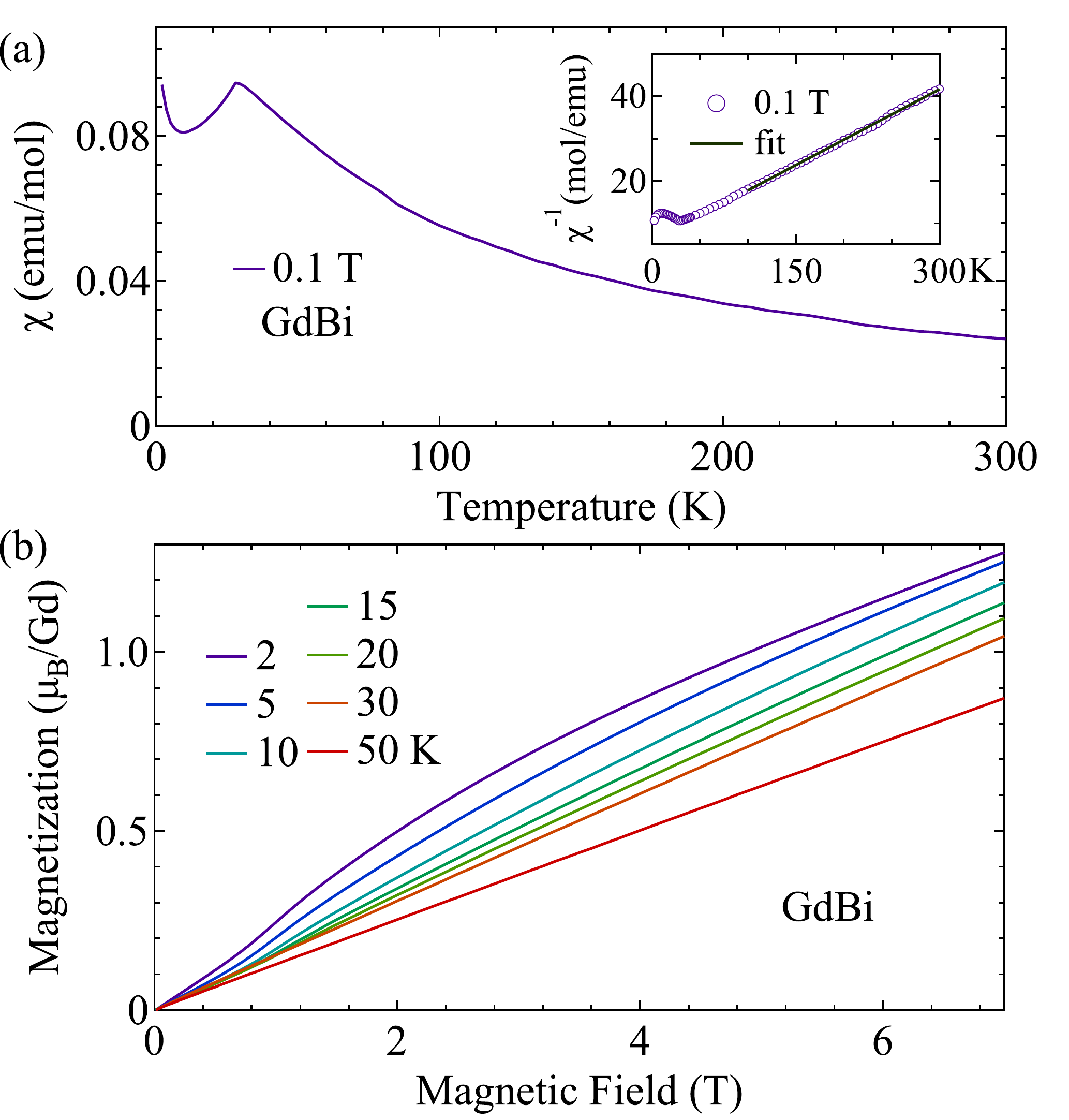}
	\caption{(a) Magnetic susceptibility $(\chi)$ as function of Temperature in $0.1$~T magnetic field, (inset) $\chi^{-1}$ fitted in Curie-Weiss law in paramagnetic region. (b) Field dependence of Magnetization at different temperature for field parallel to [100] (or equivalent) direction.}
	\label{Fig2}
\end{figure}

The temperature dependence of magnetic susceptibility $\chi(T)$ measured in applied magnetic field $(B)$ of 0.1~T, parallel to [100] direction in the $T$ range $2 - 300$~K is shown in Fig.~\ref{Fig2}(a).  At $T = 27.5$~K a sharp drop in the $\chi$ confirms the antiferromagnetic ordering.  At high temperature $\chi$ shows a clear Curie-Weiss behaviour and the $\chi (T)$ data follows the Curie-Weiss law: $\chi (T) = C/(T - \theta_p)$, where $C$ is the Curie constant and $\theta_{\rm p}$ is the paramagnetic Weiss temperature.  The effective magnetic moment $\mu_{\rm eff}$ of Gd$^{3+}$ ions can simply be obtained by the relation $\mu_{\rm eff} = \sqrt{8C}$.  The inverse $\chi (T)$  plot is shown in the inset of Fig.~\ref{Fig2}(a).  The solid line shows the fit to the Curie-Weiss law.  From the fitting we obtained $\theta_{\rm p} = -48.6$~K and the Curie constant $C = 8.02$~emu/mol.  The estimated $\mu_{\rm eff}$ from the fitting is $8.1~\mu_{\rm B}$/Gd which is nearly equal to the theoretical value of $7.9~\mu_{\rm B}$ of a free Gd$^{3+}$ ion.  The negative value of $\theta_{\rm p}$ confirms the antiferromagnetic correlations.  

The isothermal magnetization $(M)$ measured at various fixed temperatures is shown in Fig.~\ref{Fig2}(b).  The $M(B)$ curves for $T < T_{\rm N}$ shows a  small change of slope at around $1$~T signalling a subtle spin re-orientation followed by steady increase without any sign of saturation up to a magnetic field of $7$~T.  An estimation of the critical field at which the magnetization attains the saturation value for an antiferromagnet at $T = 0$~K can be estimated using the mean field model.  The expression for the critical field $H_{\rm c}$ is given by~\cite{anand2014physical}:

\begin{equation}
	\label{eqn1}
	H_{\rm c} = \frac{M_{\rm s}}{\chi( T_{\rm N})},
\end{equation}
where $M_{\rm s} = 7~\mu_{\rm B}$/Gd, the saturation magnetization  of Gd$^{3+}$ ion and $\chi(T = T_{\rm N})$ is obtained from Fig.~\ref{Fig2}(a).  Substituting the value of $\chi (T_{\rm N}) = 0.09454$~emu/mol~$= 1.6927~\times~10^{-5}~\mu_{\rm B}$~Oe$^{-1}$~Gd$^{-1}$, the critical field $H_{\rm c}$ is estimated as $41$~T.  This estimated value of $H_{\rm c}$ is in good agreement with the high field magnetization data, where the $M_{\rm sat}$ is attained at a critical field of $H_{\rm c} = 42$~T~\cite{li1997magnetic}.  

\begin{figure*}
	\includegraphics[width=0.8\linewidth]{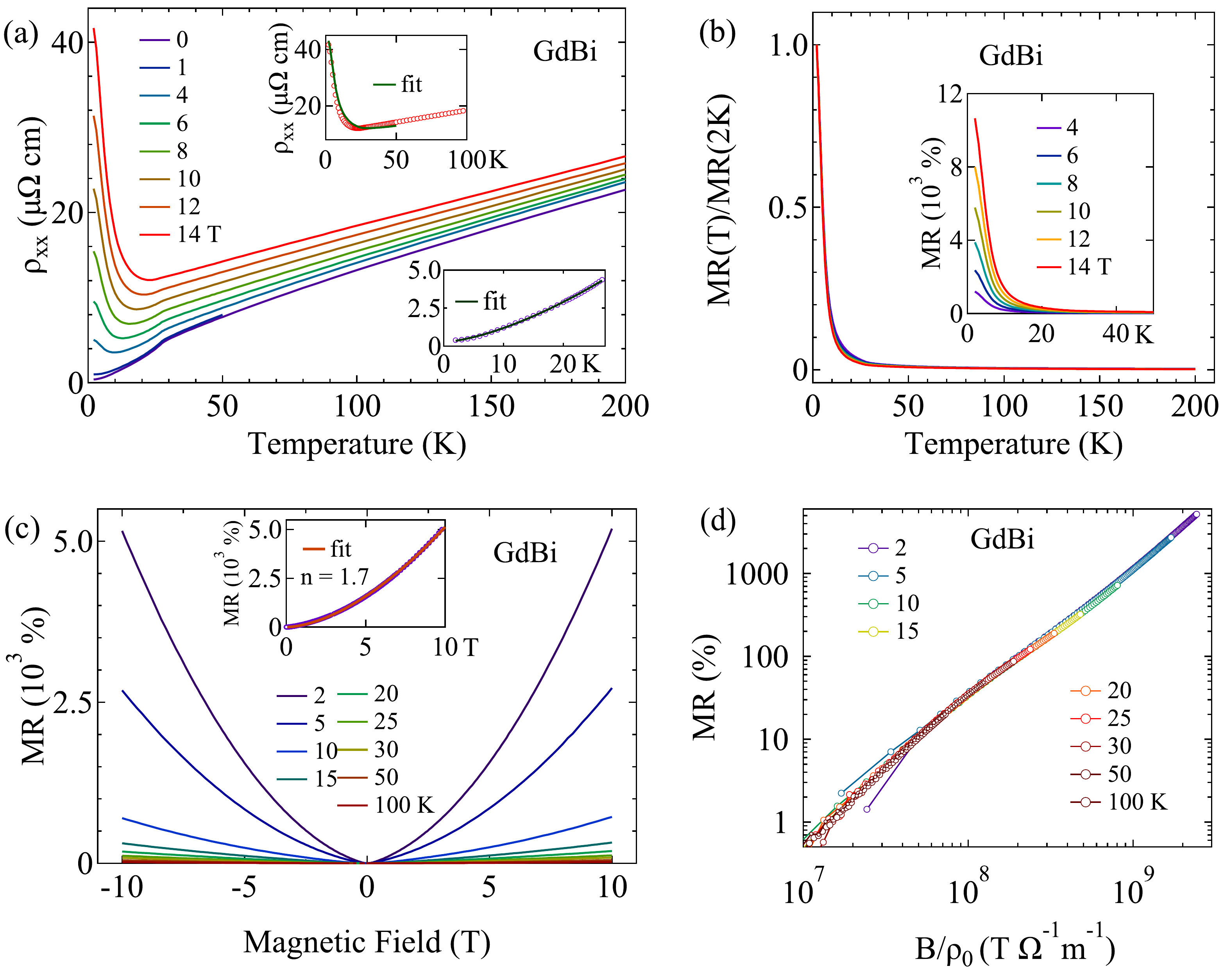}
	\caption{(a) Temperature dependence of electrical resistivity of GdBi at different applied magnetic fields, (top inset) fitting of resistivity at $14$~T magnetic field with Eq.(\ref{eq1}), (bottom inset) power law fitting of low temperature resistivity. (b) Temperature dependence of MR normalized with its value at $T2$~K at different magnetic fields. (inset) MR as function of temperature. (c) Field dependence of MR at various temperatures, (inset) power law fitting of MR at $2$~K . (d) Kohler's scaling of MR at different temperature. }
	\label{Fig3}
\end{figure*}

\subsection{Electrical Resistivity and Magnetoresistance}

The temperature dependence of electrical resistivity $\rho_{\rm xx}(T)$ measured in zero field is shown in the main panel of Fig.~\ref{Fig3}(a).  In the absence of magnetic field the $\rho_{xx}(T)$ decreases linearly as the temperature is decreased  and at $T_{\rm N} = 27.5$~K a sharp drop  in $\rho_{xx}(T)$ is observed due to the reduction in the spin disorder scattering.  Below $T_{\rm N}$, the resistivity drops rapidly and attains a value of about $390$~n$\Omega$cm at $2$~K.   The residual resistivity ratio (RRR) of the sample is estimated as 82 suggesting a high quality of crystal.  The zero field $\rho_{xx}(T)$ data, below $T_{\rm N}$  was fitted to the power law $\rho_{xx}(T) = \rho_0 + a T^n$ as shown in the inset of Fig.~\ref{Fig3}(a).  The best fit to the data was obtained for $n = 1.5$, typically for most of the rare-earth compounds the resistivity show a power law behavior, due to the $e-e$ scattering,  however in the present case the low $T$ resistivity data shows a $T^{1.5}$ behaviour. According to the spin fluctuation theory by Moriya et al.~\cite{Moriya}, the $T^{1.5}$ dependence is observed in antiferromagnetic materials near a quantum critical point~\cite{julian1998non, Moriya}.  The $T^{1.5}$ behaviour in a localized $f$ electron system GdBi warrants further investigation.

The temperature dependence of $\rho (T)$ at various applied magnetic fields is also shown in Fig.~\ref{Fig3}(a).  The overall behaviour of $\rho_{xx}(T)$ remains the same in  the paramagnetic state in applied magnetic fields.  The antiferromagnetic transition at $T_{\rm N} = 27.5$~K remains robust for fields as high as $14$~T without any shift which is substantiated with the magnetic susceptibility and heat capacity data as well.  However, $\rho_{xx}(T)$ shows an upturn well below $T_{\rm N}$ for fields greater than $1$~T.  The upturn increases more rapidly with higher magnetic fields and resembles  a metal-to-insulator like transition (MIT).  Similar behaviour has been observed in other $RBi$ (R = Pr, Ho, Er) compounds~\cite{PhysRevB.99.245131, Wu_2019, PhysRevB.102.104417}.  It is interesting to note that the $\rho_{xx}(T)$ goes through a minimum before the upturn and  this minimum shifts to higher temperature as the magnetic field is increased.  The magnetic field driven MIT has been observed in systems like WTe$_2$, NbP~\cite{Ali_2014, shekhar2015extremely} and semimetallic compounds like MoSi$_2$, WSi$_2$, WP$_2$~\cite{PhysRevB.97.205130, PhysRevB.102.115158, kumar2017extremely} .  Different mechanisms have been put forward for this type of field induced MIT and large MR in topological materials.  For example, in WTe$_2$ the extremely large MR is attributed to the  perfect electron-hole resonance, while in the case of LaBi and LaSb the large MR is attributed to the orbital texture~\cite{tafti2016temperature}.  All these mechanisms are for non-magnetic systems.  GdBi exhibits an antiferromagnetic transition and typically, in such kind of materials the positive magnetoresistance is  attributed to the suppression of the $T_{\rm N}$ due to applied magnetic fields.  However, $T_{\rm N}$ of GdBi is robust and no change in $T_{\rm N}$ has been observed for fields as high as $14$~T, furthermore the $M(B)$ data by Li $et~al.$~\cite{li1997magnetic}, have revealed that the magnetization increases linearly and attains the field induced ferromagnetic state at around $42$~T, which suggests the gradual spin re-orientation. The scattering due to this staggered moment together with near compensation of charge carriers and a very low residual resistivity (vide infra) results in such a large MR.

To understand the field dependence of the electrical resistivity further, we plotted the normalized temperature dependence of  resistivity MR($T$)/MR($2$K), measured in different fields from $4$ to $14$~T in steps of $2$~T, as shown in Fig.~\ref{Fig3}(b).  It is interesting to see that the normalized curves fall on to a single curve suggesting the $T$-dependent MR remains almost the same for all magnetic fields. Hence, it can be said that the low temperature behavior of $\rho_{xx}(T)$ is metallic rather insulating in high magnetic fields~\cite{PhysRevLett.115.046602}.  A similar behavior is observed in the magnetically ordered ErBi~\cite{PhysRevB.102.104417}. The upturn in $\rho_{xx}(T)$ at low temperature can be  well described by the Kohler's scaling rule~\cite{PhysRevB.92.180402,PhysRevB.97.235132,PhysRevB.97.245101}, according to which, the field dependent MR at different temperature will follow the same functional form MR $\propto f(B\tau)$ where $\tau$ is the relaxation time that is inversely proportional to $\rho_{0}$ as long as the scattering mechanisms at different temperatures remain same. Following the Kohler's rule, the resistivity $\rho(B,T)$ of GdBi can be written as:
\begin{equation} \label{eq1}
    \rho(B,T)=\rho_{0}(0,T)\left[1+\gamma \left(\frac{B}{\rho_{0}(0,T)}\right)^m\right]
\end{equation}
where $\rho_{0}(0,T)$ is the measured resistivity of GdBi at zero applied magnetic field. A fit of Eqn.~\ref{eq1}, keeping $\gamma$ as the only adjustable parameter, with data at $14$~T is shown in the top inset of Fig.\ref{Fig3}(a). Value of $m$ is kept fixed at 1.7, obtained from power law fitting of magnetoresistance. Deviation becomes large when the temperature is above the transition temperature $T_{\rm N}$, which maybe attributed to the change of scattering mechanism, which limits the use of Kohler's law. The above equation also shines light on the presence of a minimum in resistivity at magnetic field due to the coexistence of $\rho (0,T)$ and inverse of that term in $\rho(H,T)$.

Magnetoresistance of GdBi at different temperature is shown in Fig. \ref{Fig3}(c) as a function of magnetic field applied in transverse direction of current. At $2$~K, MR reaches $5.1\times10^3\%$ in field $10$~T and $10.9\times10^3\%$ in $14$~T (not shown here) which is extremely large. Best fit of the MR data at $2$~K with power law is shown in the inset of \ref{Fig3}(c) where the obtained exponent value is $1.7$. For a perfectly compensated semimetal one would expect, this value to be 2. In this present case the exponent value reveals that the charge carriers are nearly compensated and this has been confirmed from the Hall data (to be discussed below). Value of the exponent decreases as temperature is increased from $2$~K up to $20$~K after that it again increases till $100$~K with value 1.8, probably due to change of carrier concentration.  To verify the applicability of Eqn.\ref{eq1} to the resistivity data, we performed the Kohler's scaling rule to the field dependent magnetoresistance plot, shown in Fig. \ref{Fig3}(d), where MR is plotted against $(B/\rho_{0})$ at different temperatures, and as can be seen, all MR are collapsing onto a single curve as predicted by the scaling rule. Deviation from a straight line behaviour is mainly attributed to the different scattering mechanism in GdBi.   This also describes why the fitting shown in the top inset of Fig. \ref{Fig3}(a) using Eqn.\ref{eq1} is deviating at temperature above $T_N$.

We have used Hall  resistivity $(\rho_{xy})$ and  linear resistivity $(\rho_{xx})$ data to estimate the carrier concentration  of  GdBi. Hall measurements were performed in five-probe geometry followed by antisymmetrization of the data to minimize the contribution of linear resistivity $(\rho_{xx})$. The curved nature of Hall data implies the existence of multiple charge carriers. Two-band model is used here to fit the linear conductivity $(\sigma_{xx})$ and hall conductivity $(\sigma_{xy})$. In semi-classical two band model, complex conductivity can be written as:

\begin{equation}\label{eq2}
    \sigma=e\left[\frac{n_{e}\mu_{e}}{1+i\mu_{e}B}+\frac{n_{h}\mu_{h}}{1-i\mu_{h}B}  \right]
\end{equation}

where $e$ is the magnitude of elementary charge and $\sigma_{xx}$ and $\sigma_{xy}$ are obtained from real part and imaginary part of $\sigma$ respectively. $\sigma_{xx}$ and $\sigma_{xy}$ are calculated from experimental $\rho_{xx}$ and $\rho_{xy}$ using the following relations:

\begin{equation}\label{eq3}
    \sigma_{xx}=\frac{\rho_{xx}}{\rho_{xx}^2+\rho_{xy}^2}
\end{equation}

\begin{equation}
    \sigma_{xy}=\frac{\rho_{xy}}{\rho_{xx}^2+\rho_{xy}^2}
\end{equation}

\begin{figure}[h]
\includegraphics[width=1\linewidth]{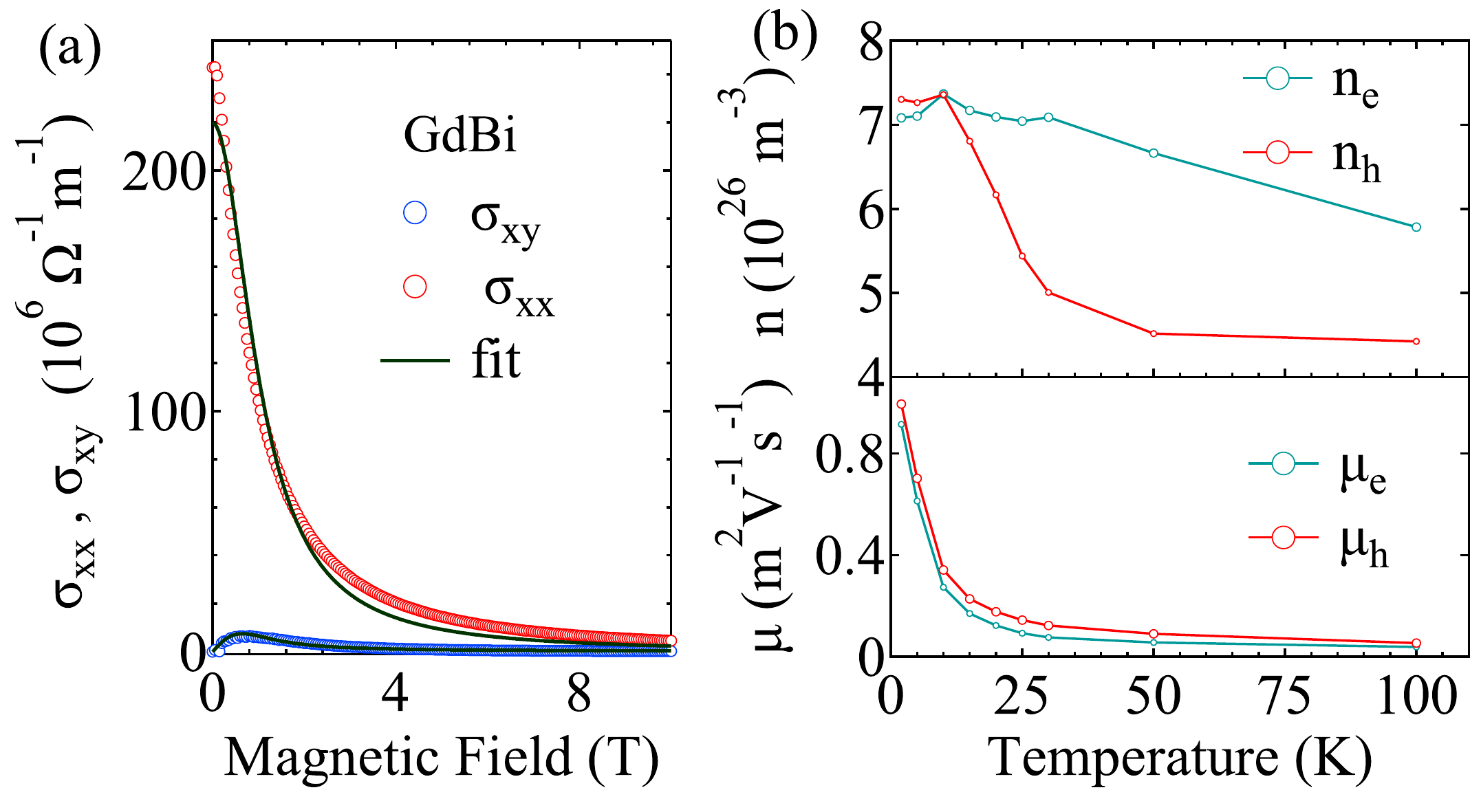}
\caption{(a) $\sigma_{xx}$ and $\sigma_{xy}$ of GdBi as a function of magnetic field. Solid lines show the fit with two band model. (b) (top) Electron and hole density, (bottom) electron and hole mobility at different temperature, obtained from two band model fitting.}
\label{Fig4}
\end{figure}
We simultaneously fitted the $\sigma_{xx}$ and $\sigma_{xy}$ data at $2$~K to the real and imaginary part of Eqn.\ref{eq2} and the best fit is shown in Fig.\ref{Fig4}(a). Calculated values of carrier density and mobility at different temperatures is shown in Fig.\ref{Fig4}(b). At $2$~K the electron  and hole density are estimated as $7.08\times10^{26}$ m$^{-3}$ and $7.30\times10^{26}$m$^{-3}$ which explains the deviation of MR  from quadratic behaviour in field. As temperature is increased, we see a crossover in electron and hole density, which is consistent with the observation that the exponent value ($n$) in power law fitting of MR is increasing towards the value 2 as temperature is increased. The estimated mobility at $2$~K is also quite large, reaching $0.914$~m$^{2}$V$^{-1}$s$^{-1}$ for electrons and  $0.994$~m$^{2}$V$^{-1}$s$^{-1}$ for holes. These values of carrier concentration and mobility are nearly the same as obtained for ErBi and HoBi compounds~\cite{Wu_2019, PhysRevB.102.104417}.  The nearly compensated nature of the charge carriers together with relatively high mobility are the reasons for the observed large MR.

\subsection{Shubnikov-de Hass (SdH) Oscillations}

\begin{table*}[t]\centering 
	\caption{Parameters estimated from SdH oscillation analysis: $m^*$,  effective mass; $T_D$, Dingle temperature; $A$, extremal area of the Fermi surface;  $k_{\rm F}$, Fermi wave vector; $v_{\rm F}$, Fermi Velocity; $\tau$, quantum relaxation time, and $\mu_{\rm Q}$, quantum mobility.}
	\begin{tabularx}{1.0\textwidth} { 
			>{\centering\arraybackslash}X
			>{\centering\arraybackslash}X 
			>{\centering\arraybackslash}X 
			>{\centering\arraybackslash}X 
			>{\centering\arraybackslash}X 
			>{\centering\arraybackslash}X 
			>{\centering\arraybackslash}X 
			>{\centering\arraybackslash}X 
			>{\centering\arraybackslash}X  } 
		\hline\hline\\[0.0001ex] 
		
		
		Fermi pocket & Frequency & $m^{*}$ & $T_{D}$ & $A$ & $k_{F}$ & $v_{F}$ & $\tau$ & $\mu_{Q}$\\
		& (T) & $(m_{e})$ & (K) & (nm$^{-2}$) & (10$^7$ cm$^{-1}$) & (10$^7$ cm/s) & (10$^{-13}$ s) & (cm$^2$/V s) \\[1ex]
		\hline\\[0.001ex] 
		$\alpha_1$ & 462  & 0.36 & 6.13 & 4.40 & 1.18 & 3.75 & 1.98 & 954.5 \\
		$\alpha_2$ & 872 & 0.35 & 14.66 & 8.30 & 1.62 & 5.38 & 0.83 & 416.3 \\
		$\alpha_3$ & 1641 & 0.72 & 5.22 & 15.63 & 2.23 & 3.56 & 2.33 & 564.1 \\
		$\alpha_4$ & 2205 & 0.72 & 5.24 & 21.01 & 2.58 & 4.13 & 2.32 & 562.4 \\
		$\alpha_5$ & 2564 & 0.78 & $\dots$ & 24.43 & 2.79 & 4.11 & $\dots$ & $\dots$ \\ [1ex] 
		\hline\hline 
	\end{tabularx}
	\label{Table1} 
\end{table*}

\begin{figure}[!]
	\includegraphics[width=1\linewidth]{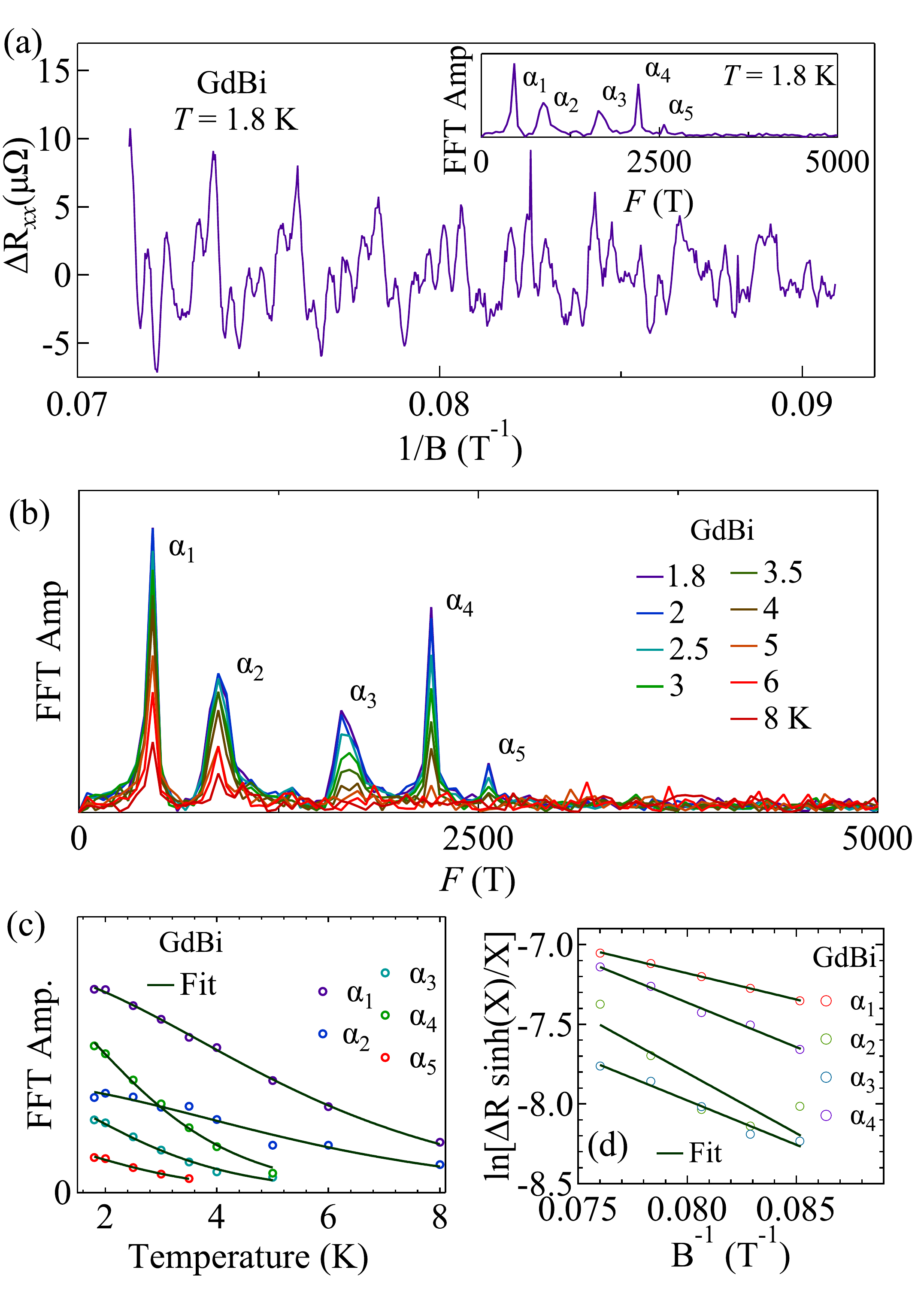}
	\caption{(a) Oscillating part of electrical resistivity of GdBi as a function of inverted magnetic field, (inset) Fast Fourier transform of oscillating part of electrical resistivity. (b) FFT amplitude at different temperature. (c) Mass plot of different frequencies. The solid lines are the fits to
		the thermal damping factor of the Lifshitz-Kosevich expression. (d) Dingle plots of different frequencies. }
	\label{Fig5}
\end{figure}

We have observed oscillation in MR when the field is ramped from $11 - 14$~T, up to $8$~K temperature beyond which it is not discernible. The background subtracted SdH oscillation ($\Delta R_{xx}$) is shown in Fig.~\ref{Fig5}(a). It is quite obvious from the shape of the oscillation that it possess multiple frequencies. A fast Fourier transform (FFT) of the oscillation revealed as many as $5$ fundamental frequencies at 461~T ($F_{\alpha_1}$), 871~T ($F_{\alpha_2}$), 1640~T ($F_{\alpha_3}$), 2203~T ($F_{\alpha_4}$) and 2562~T ($F_{\alpha_5}$), as shown in  the inset of Fig.~\ref{Fig5}(a). The frequencies may correspond to five pockets  in the  Fermi surface~\cite{Duan_2007,PhysRevB.97.081108}, two hole pockets at $\Gamma$ point and three electron pockets at $X$ points in the Brillouin zone. 
 
 The extremal cross section area ($A_{\alpha_i}$) of these Fermi pockets are calculated from the Onsager relation: $F_{\alpha_i} = (\hbar A_{\alpha_i}/2\pi e)$, see Table~\ref{Table1}. The oscillatory part $\Delta R_{xx}$ can be described by Lifshitz-Kosevitch (LK) expression~\cite{shoenberg2009magnetic}. The temperature dependent amplitude, obtained from FFT of $\Delta R_{xx}$, is shown in Fig.~\ref{Fig5}(b), that shows the oscillation amplitude of individual frequency decreases with increase in temperature, and it follows thermal damping factor $X/sinhX$, where $X=(2\pi^2m^{*}_{\alpha_i}k_BT)/(e\hbar B)$. Here, index $i$ represents i-th frequency, $m^{*}_{\alpha_i}$ is the effective mass corresponding to the frequency $F_{\alpha_i}$. Fig.~\ref{Fig5}(c) shows mass plot for different frequencies, and the extracted effective masses are given in Table~\ref{Table1}. Estimated effective mass of carriers are very similar to other rare-earth monopnictides ~\cite{PhysRevB.97.235132,PhysRevB.97.081108,PhysRevB.99.245131}. Also, the field induced oscillation amplitude damping follows $exp[-(2\pi^2m^{*}_{\alpha_i}k_BT_{D_{\alpha_i}})/(e\hbar B)]$, where $T_{D_{\alpha_i}}$ is the Dingle temperature of $F_{\alpha_i}$, and $\frac{1}{B}=\frac{1}{2}(\frac{1}{B_1}+\frac{1}{B_2})$ where $B_1$ and $B_2$ represents the range of applied magnetic field used. As multiple frequencies results to such $\Delta R_{xx}$ oscillation, FFT amplitude $(A_{\Delta R})$ of individual frequency in different field segments can give the field dependent amplitude variation. From the $ln[A_{\Delta R}sinh(X)/X]$ vs. $1/B$ fit, see Fig.~\ref{Fig5}(d), obtained Dingle temperatures are used to calculate the quantum relaxation time $\tau_{\alpha_i} (=\hbar/2\pi k_B T_{D_{\alpha_i}})$ and quantum mobility $\mu_{Q_{\alpha_i}} (=e\hbar/2\pi k_BT_{D_{\alpha_i}}m^{*}_{\alpha_i})$ of carriers for different pockets, listed in Table~\ref{Table1}. Here $F_{\alpha_5}$ is relatively large to extract the field dependent amplitude variation within applied field range. A much lower temperature and high magnetic field are necessary to investigate further.



\subsection{Specific heat}

\begin{figure}[!]
 	\includegraphics[width=1\linewidth]{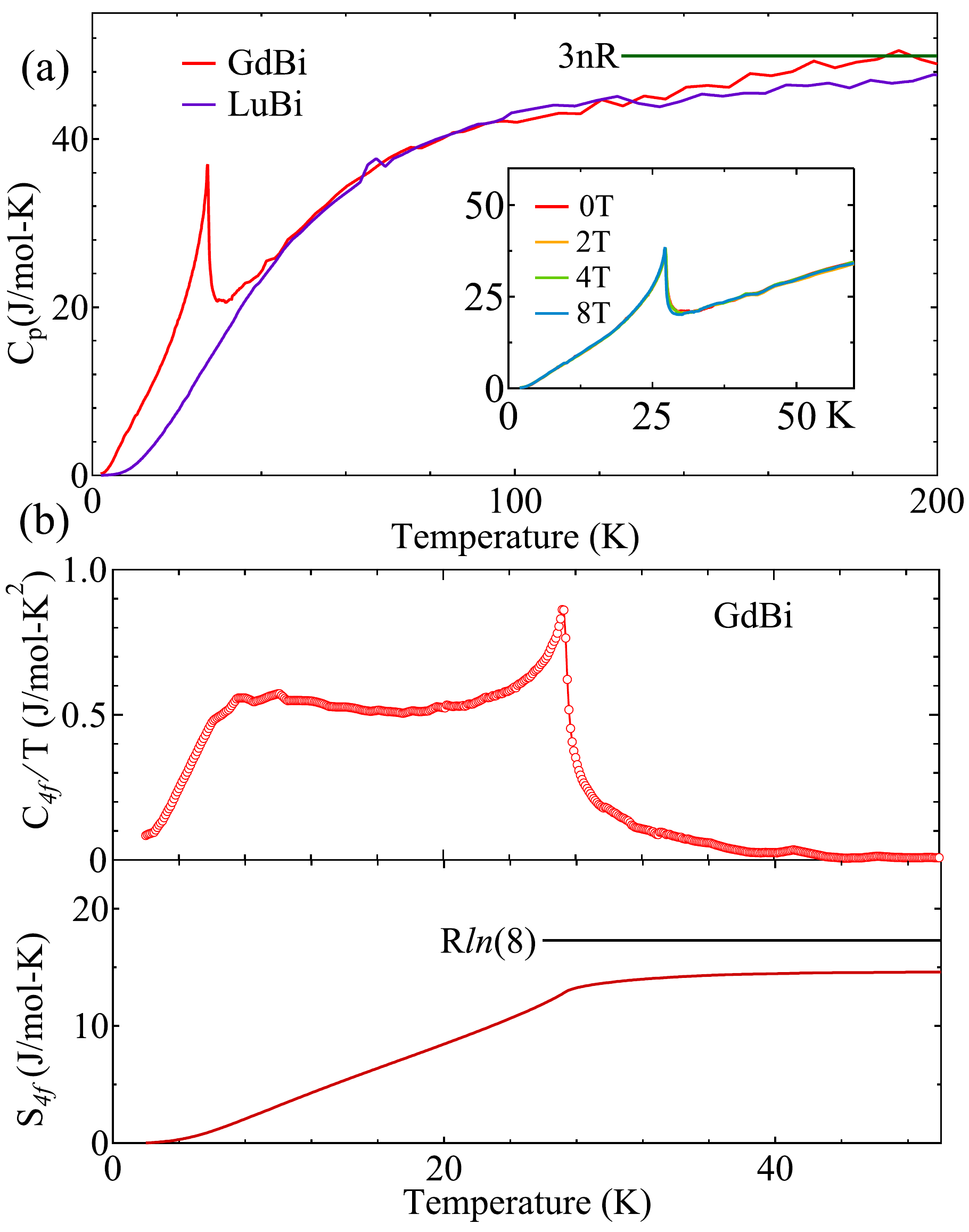}
 	\caption{(a) Specific heat $(C_p)$ of GdBi and LuBi as function of temperature, (inset) $C_p$ vs. Temperature at different magnetic field. (b) (top) $C_{4f}/T$ versus temperature  and  (bottom) magnetic entropy as a function of temperature.}
 	\label{Fig6}
 \end{figure}

 The heat capacity $(C_{p})$ of GdBi single crystal in zero magnetic field in the temperature range $2$ to $200$~K is  shown in the main panel of Fig.\ref{Fig6}(a). The sharp $\lambda$ shaped peak  at $27.5$~K is observed confirming the antiferromagnetic ordering in this compound.  The heat capacity attains the value of 49.89~J K$^{-1}$mol$^{-1}$ at $200$~K which is the expected Dulong-Petit limiting value of $3nR$.  The field dependence of heat capacity is shown in the inset of Fig.~\ref{Fig6}(a).  It is evident that the magnetic field does not have any effect on the magnetic ordering.  Typically, in antiferromagnetic materials the long range interaction competes with the applied magnetic field, resulting in the lowering of the ordering temperature.  As we have already seen in the $M(H)$ data, the $4f$ moments align to the applied field direction in an extremely slow rate and the field induced ferromagnetic state is achieved at high magnetic field of $42$~T~\cite{li1997magnetic}.  Hence fields up to $8$~T do not have any effect to the heat capacity peak.     
 
Next, we estimate the magnetic contribution $C_{4f}(T)$ to the heat capacity of GdBi, to accomplish this we have grown the single crystal of LuBi which possesses the same rock-salt type cubic crystal structure and a completely filled $4f$-shell.  Due to the difference in atomic mass of Gd and Lu, obtained $C_{p}$ of LuBi cannot completely mimic the lattice contribution of GdBi. A correction to the $C_{p}$ is required~\cite{PhysRevB.43.13137} to obtain the correct value of specific heat. The correction factor in this case is calculated to be 1.02, that is very small and here neglected. After subtracting the lattice and electronic part, obtained magnetic specific heat of GdBi is shown in Fig.\ref{Fig6}(b)(bottom) as $C_{4f}/T$.  A broad hump is observed in the  $C_{4f}/T$ plot at low temperature.  Such type of hump arises for the systems with $(2S+1)$-fold degenerate ground state with large $S$-values~\cite{johnston2011magnetic}.  For large $S$ the entropy is large, in order to accommodate the increased entropy a hump appears in the heat capacity.  For $S = 7/2$ systems according the mean field theory the hump $C_{4f}(T)$ data appears at $T~\le~T_{\rm N}/3$~\cite{johnston2011magnetic}. The hump appears at $9.2$~K, which agrees well with the MFT for $T_{\rm N} = 27.5$~K in GdBi.   The magnetic entropy $S_{4f}$ attains a value of 13.02~J K$^{-1}$mol$^{-1}$ at $T_{\rm N}$ which is $75\%$ of $R$ln$(2S+1)= 17.3$~J K$^{-1}$mol$^{-1}$ for $S = 7/2$.  The reduction in the entropy may be attributed to the inaccurate estimate of the lattice contribution to the heat capacity.

\section{Conclusion}

High quality single crystal of GdBi has been grown by high temperature solution growth.  We have performed a systematic study on the magnetic and electrical transport properties.  From the SdH quantum oscillations we have analysed the Fermi surface properties.  The magnetic measurements revealed that GdBi undergoes an antiferromagnetic transition at $27.5$~K.  The isothermal magnetization $M(H)$ did not show any sign of saturation and reached value of about $1.25$~$\mu_{\rm B}$/Gd  at $7$~T, where as the saturation moment is $7$~$\mu_{\rm B}$/Gd.  The electrical resistivity confirmed the antiferromagnetic ordering by displaying a sharp drop in the resistivity at $27.5$~K and the overall resistivity behaviour was typical metallic like.  With application of magnetic field the electrical resistivity displayed a huge upturn in the magnetically ordered state at low temperature as observed in most of the compensated semimetallic systems.  The MR also displayed a large value of the order 10$^4$\% without any sign of saturation.  Hall effect studies revealed multiple type of charge carriers and a near compensation of the  charge carriers.  From the SdH oscillation studies we have estimated the effective masses of the observed five different frequencies which are almost of the same order as observed in other rare-earth monopnictides.  A high field MR measurements at low temperatures and fields greater than $14$~T will throw more light on the observed non-saturating magnetoresistance.  

%
\end{document}